\documentclass[preprint,12pt]{elsarticle}

\usepackage{amssymb}
\usepackage{amsmath}

\journal{Nuclear Physics B}

\begin{document}

\begin{frontmatter}



\title{Spatio-temporal Dynamical Indices for Complex Systems}

\author[1]{Chenyu Dong}
\author[2,3,4]{Gabriele Messori}
\author[5,6,7]{Davide Faranda}
\author[8,9]{Adriano Gualandi}
\author[10]{Valerio Lucarini}
\author[1]{Gianmarco Mengaldo\corref{cor1}}
\cortext[cor1]{Corresponding author. Email: mpegim@nus.edu.sg}

\affiliation[1]{organization={National University of Singapore},
            country={Singapore}}

\affiliation[2]{organization={Department of Earth Sciences, Uppsala University},
            country={Sweden}}

\affiliation[3]{organization={Swedish centre for impacts of climate extremes (climes), Uppsala University},
            country={Sweden}}

\affiliation[4]{organization={Department of Meteorology and Bolin Centre for Climate Research, Stockholm University},
            country={Sweden}}

\affiliation[5]{organization={Laboratoire des Sciences du Climat et de l'Environnement, UMR 8212 CEA-CNRS-UVSQ, Universit\'e Paris-Saclay \& IPSL, CEA Saclay l'Orme des Merisiers},
            city={Gif-sur-Yvette},
            country={France}}

\affiliation[6]{organization={Laboratoire de Météorologie Dynamique/IPSL, École Normale Supérieure, PSL Research University, Sorbonne Université, École Polytechnique, IP Paris, CNRS},
            city={Paris},
            country={France}}
            
\affiliation[7]{organization={London Mathematical Laboratory},
            city={London},
            country={UK}}

\affiliation[8]{organization={University of Cambridge},
            city={Cambridge},
            country={UK}}

\affiliation[9]{organization={Istituto Nazionale di Geofisica e Vulcanologia},
            country={Italy}}

\affiliation[10]{organization={School of Computing and Mathematical Sciences, University of Leicester},
            country={UK}}

\begin{abstract}
Complex systems span multiple spatial and temporal scales, making their dynamics challenging to understand and predict. 
This challenge is especially daunting when one wants to study localized and/or rare events. 
Advances in dynamical systems theory, including the development of state-dependent dynamical indices, namely local dimension and persistence, have provided powerful tools for studying these phenomena. 
However, existing applications of such indices rely on a predefined and fixed spatial domain, that provides a single scalar quantity for the entire region of interest.
This aspect prevents understanding the spatially localized dynamical behavior of the system.  
In this work, we introduce Spatio-temporal Dynamical Indices (SDIs), that leverage the existing framework of state-dependent local dimension and persistence.
SDIs are obtained via a sliding window approach, enabling the exploration of space-dependent properties in spatio-temporal data.
As an example, we show that, through this framework, we are able to reconcile previously different perspectives on European summertime heatwaves.
This result showcases the importance of accounting for spatial scales when performing scale-dependent dynamical analyses.
\end{abstract}







\end{frontmatter}



\section{Introduction}
\label{sec:introduction}

Since Lorenz's pioneering work~\cite{lorenz1963deterministic}, dynamical systems theory has served as a powerful framework for studying complex systems. 
These systems are inherently nonlinear, high-dimensional, and multiscale, which makes them difficult to characterize and capable of exhibiting a wide range of dynamical behaviors~\cite{altman2018curse,gao2024similarity,gao2025inhomogeneity}.
While much of the existing literature has focused on the global (or average) properties of these systems~\cite{grassberger1986climatic,lorenz1991dimension}, such approaches may not be directly applicable to the study of extreme events, whose onset mechanisms are often linked to a system's instantaneous properties.
Recent advances in dynamical systems theory have provided a mathematically rigorous, and purely data-driven framework for analyzing local, instantaneous, state-dependent dynamical properties of complex systems. 
This data-driven framework is achieved by looking at the statistics of close recurrences of the orbit with respect to a reference state of interest ${\boldsymbol{\zeta}}$, thereby inferring its dynamical properties~\cite{lucarini2016extremes,faranda2017dynamical}. 
This framework provides two metrics: (i) the local dimension $d$, which is the local geometric information about the system's complexity, and (ii) the clustering of recurrences around a state, $\theta$, which is the reciprocal of the local persistence time $\Theta = \theta^{-1}$.
Indeed, these two indices have been applied to several problems, providing insightful analysis for temperature extremes and their atmospheric drivers~\cite{hochman2022dynamics,holmberg2023link,messori2017dynamical}, the life cycles and transitions of weather regimes~\cite{hochman2021atlantic,lee2024dynamical,platzer2021dynamical}, ocean variability~\cite{liu2021dynamical} and earthquake dynamics~\cite{gualandi2024similarities}. 

%
%
The works just mentioned address the different phenomena they target looking at the system's dynamical behavior on a single and fixed spatial domain that is of interest to the specific study. 
This implies that the indices adopted for each study only `see' one single spatial scale, an approach that does not fully capture the possibly spatially-varying dynamical nature of several of these and other real-world systems.

Indeed, many geophysical and other spatio-temporal systems exhibit dynamics that span a wide range of scales, both spatially and temporally; choosing a fixed region (i.e., spatial domain) might not fully account for these multiscale physical processes. 
In addition, pre-selecting a fixed spatial domain requires prior knowledge of the studied system and its dynamics, and may fail to account for dynamical processes originating in regions afar from the fixed domain selected (e.g., teleconnections in the context of atmospheric dynamics)~\cite{springer2024unsupervised}. 
%
%
In fact, the limitations of the fixed-domain approach just outlined were also noticed by Faranda et al.~\cite{faranda2017dynamical} in the study of tropical cyclones, where a Lagrangian perspective was proposed to better capture their dynamics. 
To address similar challenges, empirical mode decomposition (EMD) was recently applied to decompose a given system into components at different temporal scales, prior to calculating the local indices $d$ and $\theta$~\cite{alberti2021small,alberti2023scale}. 
This approach reveals both the systems' scale-dependent dynamical characteristics and the interactions between different scales~\cite{alberti2021small}. 
However, when applied to high-dimensional spatio-temporal data, the use of EMD becomes problematic, as the method does not scale efficiently and its decomposition may suffer from robustness issues in multidimensional settings~\cite{lv2016multivariate}.
Another study on the dynamics of slow earthquakes divided the spatio-temporal system into multiple subsections based on geographical locations, with the dynamical properties of each subsection characterized individually. 
This type of strategy can be adopted only if a priori domain-specific knowledge is available.
The two examples above show that methods to characterize the multiscale nature of complex systems frequently rely on explicit decompositions or domain-specific assumptions. 
These examples highlight the need for a new framework that can characterize scale-dependent dynamical properties bypassing the limitations just outlined.

%
%
In this work, we address the above-mentioned limitations by introducing Spatio-temporal Dynamical Indices (SDIs), which extend the framework of local dynamical indices through the incorporation of a spatial sliding window approach.
This enables a more flexible analysis of the dynamical properties of complex systems, as it no longer relies on a pre-defined domain and can be applied to uncover distinct dynamical behaviors across the spatial domain of interest.
As a methodological paper, we introduce this novel approach in detail and demonstrate its effectiveness through real-world complex problem. 
To this end, we apply it to European summertime heatwaves -- high-impact events that have been extensively studied in the literature -- thus providing an ideal testbed for our method~\cite{slonosky2002does,miralles2014mega,zschenderlein2019processes,russo2015top,stefanon2012heatwave}.
In particular, we demonstrate that our approach can reconcile two seemingly different perspectives on heatwaves — one rooted in dynamical systems theory and the other in meteorology.
We note that although the scale-dependent nature of the dynamics motivates our approach, we do not systematically investigate this aspect through the use of multiple sliding windows with continuously varying sizes in this work. 
Instead, we base our analysis on a pre-selected window size and compare the results with those obtained over a large-scale domain.
Nonetheless, we emphasize the potential of this framework to be applied across multiple spatial scales in order to obtain a more comprehensive view of the scale-dependent dynamical properties and cross-scale interactions of complex systems.

\section{A new approach to reveal scale-dependent dynamics}
\label{sec:methods}
%
%
Grounded at the intersection of dynamical systems theory and extreme value theory, the two aforementioned dynamical indices, local dimension ($d$) and inverse persistence ($\theta$) can be computed based on the recurrences of $\zeta$ (i.e., when the system visits states neighboring $\zeta$ in the phase space). 
The local dimension $d$ can be considered a proxy for the effective number of active degrees of freedom of the system locally around $\zeta$. 
It therefore reflects the complexity of the dynamics near the state $\zeta$; if a system has more active degrees of freedom, its complexity is greater. 
The inverse persistence ($\theta$) is derived from the extremal index, which is defined in extreme value theory to measure the clustering of extremes. 
For a given state $\zeta$, $\theta$ can be interpreted as a measure of the inverse of the system's persistence time near that state and is defined between 0 and 1. 
A high value of $\theta$ indicates that the system will quickly leave the neighborhood of $\zeta$, while a value of $\theta$ close to 0 suggests that the system will persist in states resembling $\zeta$. 
Both $d$ and $\theta$ can be related to the intrinsic predictability of the system at a specific time~\cite{hochman2019new}. 
States with high local dimension ($d$) and high inverse persistence ($\theta$) are usually considered less predictable due to their complex and fast dynamics. 
This has been confirmed for real-world datasets using a new instantaneous index that directly measures predictability within the same theoretical framework~\cite{dong2024revisiting}. 
A more detailed description of these indices can be found in the Appendix A, with the full mathematical derivation provided in Faranda et al.~\cite{faranda2023statistical}. 

When applying these dynamical indices to real-world complex systems, it is generally impossible to access all variables to construct their full phase space. Instead, we can focus on a subset of available observables (e.g., SLP for characterizing  atmospheric circulation) to construct an approximated phase space.
Although this is not a strict phase space in the formal sense, studying the dynamics of subsystems within this approximated phase space utilizing dynamical systems theory remains highly meaningful.

\begin{figure}[h!]
\centering
\noindent\includegraphics[width=1\textwidth]{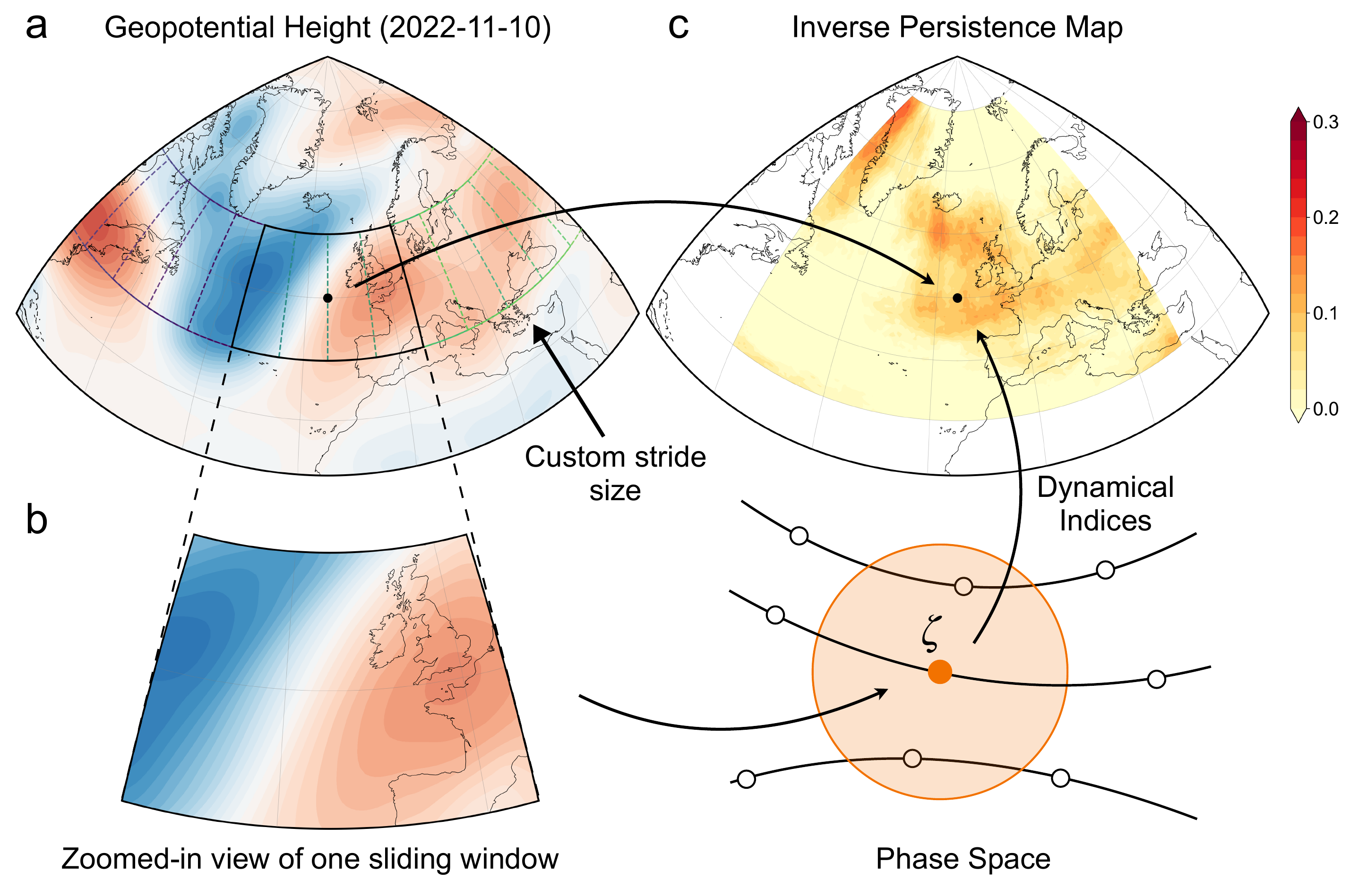}
\caption{\textbf{Schematic illustration of the computational framework for scale-dependent dynamical indices.} (a) An example of the 500hPa geopotential height map from 2022/11/10, with bounding boxes showing sliding windows spaced at uniform stride size. (b) Zoomed-in view of a sliding window and its associated idealized schematic phase space illustration. The orange dot $\zeta$ represents the state in the specific sliding window and the orange circle corresponds to neighborhood of $\zeta$ in the phase space, which is used to define recurrences and compute dynamical indices (see Materials and Methods). (c) The map of inverse persistence $\theta$. The edges of the figure are left blank because the center of the sliding windows do not extend to those regions.} 
\label{fig:fig1}
\end{figure}
%

In this study, we propose spatio-temporal dynamical indices (SDIs) that compute dynamical indices using a moving window scheme, as depicted in Fig.~\ref{fig:fig1}.
In previous studies, the evolution of a variable within a fixed spatial domain (e.g., geopotential at 500 hPa, as shown in Fig.~\ref{fig:fig1}) has been represented as a trajectory in the approximated phase space.
In contrast, we apply a sliding window with certain size across the entire domain, treating each window separately (Fig.~\ref{fig:fig1}a).
This means that we consider the atmospheric dynamics within each window as a sub-system, for which we can construct the corresponding approximated phase space and compute its instantaneous dynamical indices (Fig.~\ref{fig:fig1}b).
Therefore, instead of obtaining one numerical value for $d$ and one for $\theta$ for the entire domain, we obtain numerical values for each sliding window map for both $d$ and $\theta$. 
By assigning these values to the center of the sliding window map, and considering sliding windows covering the whole region of interest, we obtain spatial maps of the dynamical indices (Fig.~\ref{fig:fig1}c).  
This makes the dynamical indices not only `local' in the phase space (i.e., instantaneous in time) but also spatially (i.e., geographically) `local', features that can provide valuable insights.
Our use of the Euro-Atlantic region here is for illustration purposes only -- our method can be extended to the entire globe, enabling the study of e.g., teleconnections in the context of weather and climate applications. 

In Fig.~\ref{fig:fig1}, we use a window size of 40° longitude by 20° latitude as an example, while emphasizing that the window size can be customized to capture dynamics at different scales of interest.
We note that the size range of this sliding window depends on the system under study and the length of the dataset.
In the case of atmospheric systems, it must be neither too small, as this would make these sub-systems overly sensitive to boundary conditions, leading to the identification of false neighbors, nor too large, which would hinder the detection of close recurrences.

Unike approaches that rely on a fixed domain~\cite{faranda2017dynamical}, our spatial maps of dynamical indices capture the scale-dependent dynamical properties of complex systems, enabling a more detailed analysis of how dynamical behaviors change across spatial scales.
Meanwhile, compared to previous studies that employed dynamical indices based on a Lagrangian perspective to investigate hurricanes~\cite{faranda2023dynamical}, our approach can be considered a more systematic extension, with their results being a subset of ours.
It is also worth noting that our framework shares structural and objective similarities with convolutional kernels from computer vision~\cite{o2015introduction}, a cornerstone of the field that has proven highly effective.
Our method similarly employs a sliding window technique to extract scale-dependent information from data, but its foundation in dynamical systems theory enables the derivation of physically interpretable spatial maps rather than features lacking clear physical meaning.

\section{A Real-World Case Study}

To demonstrate the advantages of our approach, we focus on European summertime heatwaves as a case study.
In this section, we briefly introduce the background and relevant research of these phenomena, as well as the data used in this study.

\subsection{European Summertime Heatwaves}

European summertime heatwaves can pose a substantial threat to society and ecosystems~\cite{garcia2010review,gallo2024heat}, and it is critical to enhance our understanding of their dynamics for improving prediction capabilities and mitigating their impacts~\cite{ebi2021extreme,domeisen2023prediction}. 
These phenomena are driven by different processes and show significant inter-regional differences~\cite{zschenderlein2019processes}. 
In higher-latitude regions, European summer heatwaves are often associated with atmospheric blocking, which results in adiabatic warming from subsidence and clear-sky conditions~\cite{domeisen2023prediction,pfahl2012quantifying,stefanon2012heatwave}. 
At lower latitudes, European summer heatwaves are linked to persistent subtropical ridges, a configuration that weakens zonal flow and strengthens meridional flow, facilitating the southerly advection of hot air into southern Europe~\cite{cassou2005tropical}.
From a traditional meteorological dynamics perspective, both blocking and subtropical ridges are highly persistent atmospheric configurations~\cite{legras1985persistent}. 
However, recent studies based on dynamical systems theory have only found a weak to moderate connection between anomalously persistent circulation patterns in the mid-troposphere and summertime heatwaves, and few significant persistence anomalies of the surface circulation patterns~\cite{holmberg2023link}. 
Meanwhile, earlier studies based on dynamical indices~\cite{faranda2017dynamical,faranda2016switching}, along with other research~\cite{lucarini2020new,schubert2016dynamical}, have argued that blocking is long-lived yet transient feature of the atmospheric system, characterized by anomalously high instability and low predictability.
These conclusions differ markedly from the traditional meteorological perspective of blocking as a persistent atmospheric feature and call for a reconciliation of these divergent views. 

In section~\ref{sec:results}, we show how our new approach can address these divergent views and enrich our dynamical understanding of these phenomena.

\subsection{Data}

In this study, we used state-of-the-art ERA5 reanalysis daily mean data from 1979 to 2022~\cite{hersbach2020era5}. 
The scale-dependent dynamical indices analysis is performed on both sea level pressure (SLP) and 500 hPa geopotential height (Z500), two variables extensively used to analyze circulation patterns associated with temperature extremes~\cite{holmberg2023link,jezequel2018role,cassou2016disruption,Galfi2021,Lucarini2023}.
We remark that whilst SLP is most immediately related to describing the surface conditions of the atmosphere, the Z500 field distills the key dynamical information for the atmospheric variability of the mid-latitude because it is arguably the optimal two-dimensional field for capturing the quasi-geostrophic dynamical processes~\cite{Speranza1983,DellAquila2005,Lucarini2007}.
We have computed the dynamical indices for the entire Northern Hemisphere (180°W-180°E, 0° to 90°N), with the data downsampled from the original resolution of 0.25° to 0.5° to reduce computational costs. 
The stride size for the sliding window is set to 0.5° in both the longitude and latitude directions. 
In the main text, we present results using a window size of 40° longitude by 20° latitude, while the results for a smaller window size (20° longitude by 10° latitude) are provided in the Supplementary Information.
This window size is selected as a representative scale for our analysis, as it approximately corresponds to the typical spatial extent of cyclones and anticyclones, which can be viewed as fundamental building blocks of weather regimes in the Euro-Atlantic sector~\cite{fery2022learning}.
We deseasonalize and standardize both dynamical indices to focus our interpretation on anomalies relative to their respective reference values.

The seasonal cycle is calculated as the mean of a 31-day moving window centered on each calendar day, while standardization is performed on a grid-point-by-grid-point basis. 
For all mean values of standardized anomalies presented in this paper, we assess statistical significance using the 95\% confidence interval generated by a bootstrap test with 1,000 resamples (with replacement). To control the multiple-testing issue in spatial maps, we apply the Benjamini-Hochberg procedure~\cite{wilks2016stippling}.

We define regional summertime heatwaves over the British Isles, Scandinavia, central Europe (Germany), Iberia, and the central Mediterranean during the extended summer season (JJAS), following the bounding boxes used in Zschenderlein et al.~\cite{zschenderlein2019processes}. 
We then identify heatwave occurrences using the procedure outlined in Russo et al.~\cite{russo2015top}, with the detailed steps provided in Supplementary Information.

\section{Results} 
\label{sec:results}

\subsection{Scale-dependent circulation dynamics during heatwaves}

\begin{figure}[h!]
\centering
\noindent\includegraphics[width=1\textwidth]{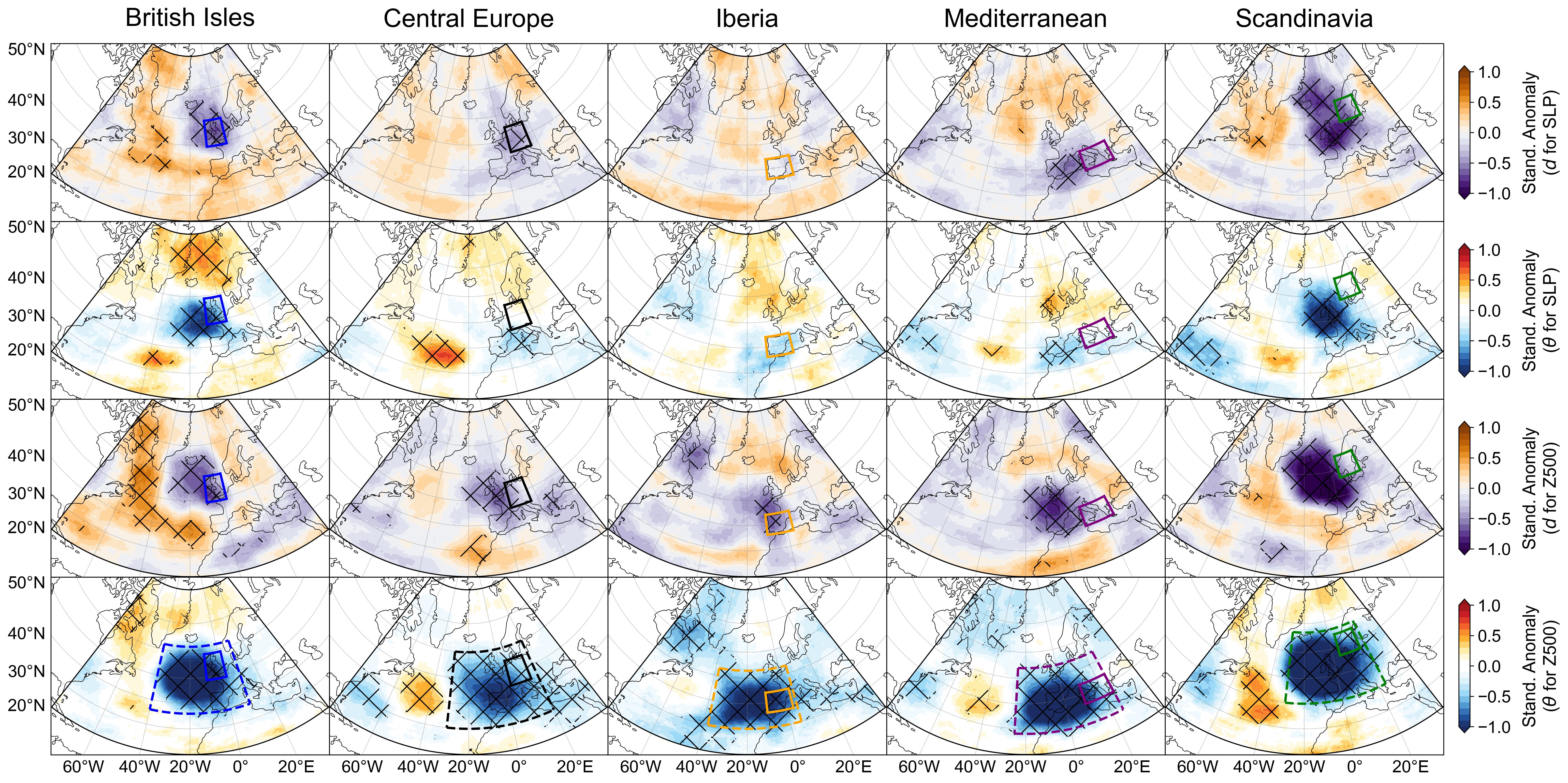}
\caption{\textbf{Composite of standardized anomalies of dynamical indices for the onset days of European summertime heatwaves.} Composite of standardized anomalies for the local dimension $d$ (rows 1 and 3) and the inverse persistence $\theta$ (rows 2 and 4) that were computed using both SLP (rows 1 and 2) and Z500 (rows 3 and 4). Each column represents heatwaves over a specific region, with the affected area indicated by a colored solid bounding box. The dashed bounding boxes in the last row highlight regions utilized in the analysis presented in Fig.~\ref{fig:fig5}. The regions studied are as follows: British Isles (blue; solid: 49°N–59°N, 10°W–2°E; dashed: 36°N–62°N, 40°W–10°E), Central Europe (black; solid: 45°N–55°N, 4°E–16°E; dashed: 30°N–60°N, 30°W–20°E), Iberian Peninsula (orange; solid: 36°N–44°N, 10°W–3°E; dashed: 30°N–53°N, 38°W–5°E), Mediterranean (purple; solid: 36°N–44°N, 10°E–25°E; dashed: 28°N–54°N, 25°W–25°E), and Scandinavia (green; solid: 57°N–65°N, 25°E–20°E; dashed: 40°N–67°N, 30°W–24°E). Cross-hatching indicate statistical significance.}
\label{fig:fig2}
\end{figure}

In Fig.~\ref{fig:fig2}, we present the composite maps of the standardized anomalies of dynamical indices for the onset days of European summertime heatwaves (we will use heatwaves hereafter, for the sake of brevity, when referring to European summer heatwaves). 
We computed the dynamical indices using SLP (Fig.~\ref{fig:fig2}, rows 1 and 2) and Z500 (Fig.~\ref{fig:fig2}, rows 3 and 4) separately to show the distinct dynamical behavior of the surface-level and mid-level troposphere.
The composite maps of the standardized anomalies of the dynamical indices derived from the two observables reveal clear differences in both magnitude and spatial configuration, with the Z500-based indices generally showing stronger anomalies. 

In the mid-level troposphere, significant negative $\theta$ anomalies (Fig.~\ref{fig:fig2}, row 4) are observed for Z500 slightly west of the target regions. 
This suggests that during the onset days of heatwaves, the upstream atmospheric configuration of the affected regions is unusually persistent, but other regions in the Euro-Atlantic sector do not have increased persistence.
In addition, the local dimension ($d$) of Z500 (Fig.~\ref{fig:fig1}, row 3) shows some negative anomalies that are co-located with, but weaker than, the negative anomalies of inverse persistence ($\theta$). 
Other regions display relatively weak or near-zero anomalies, except for British Isles heatwaves, that exhibit a strong positive anomaly over the Atlantic.
Such results suggest that Z500 patterns display reduced dynamical complexity during heatwave onset days near the affected regions.
The results for $\theta$ and $d$ on Z500 together support the view that European summertime heatwaves are accompanied by persistent and low-dimensional atmospheric configurations in the mid-troposphere from a dynamical systems perspective. 
We note that the results presented in Fig.~\ref{fig:fig2} are robust with respect to the sliding window size, as similar spatial anomaly patterns are observed when a smaller sliding window is applied (see Fig.~\ref{fig:smallscale}).

%
%

At the surface level, we note that the standardized anomalies are less pronounced, yet still reveal some coherent patterns (Fig.~\ref{fig:fig2}).
For heatwaves occurring outside Iberia, we observe significant negative anomalies in both local dimension ($d$) and inverse persistence ($\theta$) near the affected regions, similar to those in the mid-level atmosphere, although much weaker.
In Iberia, no such significant anomalies are observed for $\theta$ over the entire domain.
It is particularly noteworthy to observe distinct dynamical properties at the surface and mid-level troposphere, as also noted by Holmberg et al.~\cite{holmberg2023link}, especially since these differences are not evident in the composite anomaly maps of the raw variables (see Fig.~\ref{fig:rawvariables}). 
Such contrasts are consistent with meteorological perspectives, as surface and mid-tropospheric dynamics can differ substantially, with many features relevant to surface weather manifesting at mid-levels. 
This further illustrates that the SDI framework, through its broad applicability and sensitivity to variable choice, can reveal distinct and complementary dynamical information, thereby serving as a useful diagnostic tool for complex systems.

Considering the importance of blocking and the non-systematic link between heatwaves and blocking, we further extend our analysis to blocking events to enhance the completeness of this study, particularly given previously noted contrasting perspectives on their dynamical properties. 
From a meteorological perspective, blocking events are persistent atmospheric configurations by definition, blocking events are by definition persistent atmospheric configurations as their identification involves applying a minimum duration threshold to a particular index~\cite{pfahl2012quantifying,woollings2018blocking,kautz2021atmospheric}.
However, recent studies grounded in dynamical systems theory propose a differing view, suggesting that blocking regimes are not inherently persistent atmospheric configurations~\cite{faranda2017dynamical, hochman2021atlantic}.

In order to reconcile these views, we apply a blocking detection algorithm to examine the dynamical properties of blocking~\cite{brunner2018dependence}, with detailed steps provided in Appendix C. 
We find that, regardless of their location, blocking events are associated with geographically localized more persistent (low $\theta$) and less complex (low $d$) mid-level tropospheric flows -- see Fig.~\ref{fig:blocking}, where the second and third rows represent the anomalies of $d$ and $\theta$ for the blocking events identified by the blocking algorithm adopted.
Yet, regions surrounding the blocking center are less persistent (high $\theta$) and more complex (high $d$), and may be less predictable as also shown in operational forecasts~\cite{kautz2021atmospheric} and in agreement with the theoretical arguments set in Schubert et al~\cite{schubert2016dynamical}. 
This means while predictability increases within the blocking region, this improvement is more than offset by reduced predictability in the surrounding areas, leading to reduced predictability from a large-scale perspective.
This further analysis of blocking bridges the two seemingly different views that were present in the literature, thereby showing how the new approach can capture relevant informative features of atmospheric dynamics.

\begin{figure}[h!]
\centering
\noindent\includegraphics[width=1\textwidth]{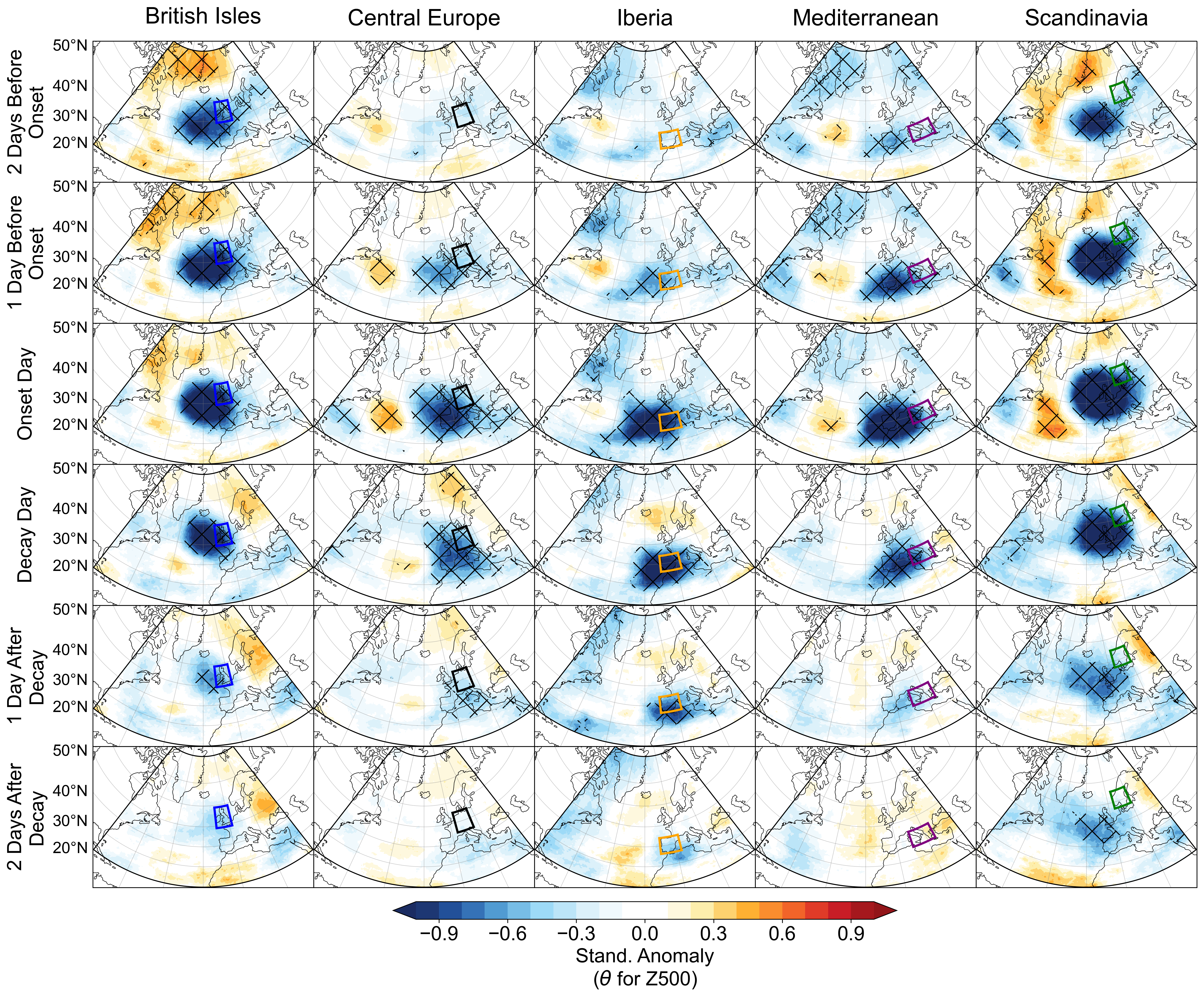}
\caption{\textbf{Temporal evolution of the composite anomaly of Z500-based inverse persistence ($\theta$) before the onset and after the decay of European summertime heatwaves.} Composite anomaly of Z500-based inverse persistence from two days before the onset to the onset day, and from the last day to two days after.}
\label{fig:fig3}
\end{figure}

A further fascinating aspect emerges when looking at the temporal coherence of the anomaly fields of the dynamical indicators as we approach the onset of the heatwaves and as we observe its decay.
In this context, Figure~\ref{fig:fig3} presents the composite anomaly maps of the Z500-based inverse persistence field. 
This variable was chosen for its strong dynamical relevance to heatwaves, as demonstrated by the results in Fig.~\ref{fig:fig2}, while the evolution of the Z500-based local dimension is shown in Fig.~\ref{fig:d_evolution}.
We observed that the onset of heatwaves exhibits a higher degree of coherence than the decay process, despite some variability across affected regions.
Previous studies, based on large deviation theory, have argued that persistent extreme events such as heatwaves and cold spells exhibit dynamical typicality~\cite{Galfi2021,Lucarini2023}.
This means that the occurrence of the extreme event requires a very special large-scale configuration of the atmospheric fields, which to a very good degree of approximation repeats itself, thus defining a class of analogues, for different individual occurrences of the event.
In other terms, such a configuration is atypical with respect to the overall statistics of the fields, but becomes typical when we focus on the days associated with the heatwave of interest. 
Indeed, our findings provide good evidence for the existence of a \textit{highway} in phase space leading to the events; see discussion in Galfi et al.~\cite{Galfi2021RNV} and first original contribution is this direction by Dematteis et al.~\cite{Dematteis2018}. 
Such coherence is largely lost after the decay of the heatwaves, because the lysis process, as opposed to the onset, does not follow a preferential path.
Our analysis could also be used to identify small-scale dynamical precursors of heatwaves -- or other extreme events in general. 
However, this would require a more detailed clustering or case study of these events, accounting for the diverse dynamical pathways leading to them~\cite{zschenderlein2019processes}, which we consider beyond the scope of this methodological paper.

\subsection{A richer dynamical view}

\begin{figure}[h!]
\centering
\noindent\includegraphics[width=0.95\textwidth]{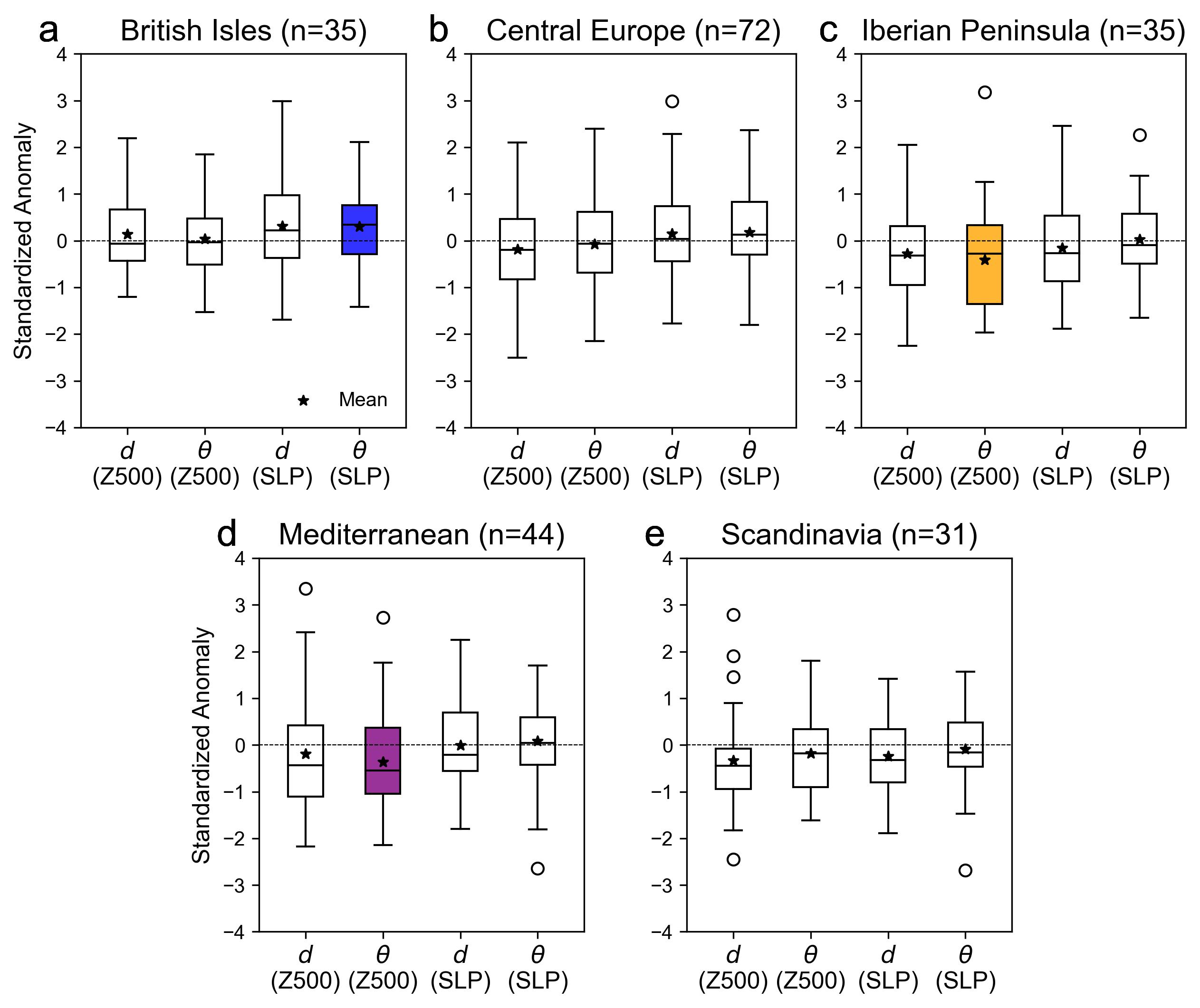}
\caption{\textbf{Box plots of standardized anomalies of dynamical indices for the Euro-Atlantic sector using the single-domain approach.} The dynamical indices are computed within the large-scale bounding box (20°N–80°N, 80°W–40°E), followed by standardization steps as described above. Each subplot shows the results for one studied region, with statistically significant results highlighted in the color representing that region, as in Fig.~\ref{fig:fig1}.}
\label{fig:fig4}
\end{figure}

To further complement our analysis with the new framework, we conducted a comparative study following the previous single-domain approach -- treating the entire Euro-Atlantic sector as a single dynamical system. 
Fig.~\ref{fig:fig4} presents box plots of standardized anomalies of dynamical indices for the entire Euro-Atlantic domain during the onset days of summertime heatwaves, with each subplot corresponding to a specific column (i.e., affected region considered) in Fig.~\ref{fig:fig2}. 
Unlike our new approach in Fig.~\ref{fig:fig2}, that shows significant anomalies for both $d$ and $\theta$ in all regions considered, the single-domain approach shows that significant results are achieved in only three cases. 
This suggests that atmospheric patterns associated with heatwaves fit a range of different dynamical properties, which are difficult to capture via the single-domain approach.
Specifically, using the single-domain method, at the mid-level, both the British Isles and Central Europe exhibit near-zero standardized anomalies, whereas the other three regions display negative anomalies for both indices.
However, only the inverse persistence ($\theta$) for the Iberian Peninsula and the Mediterranean is statistically significant.
At the surface level, no significant results are observed, except for the SLP-based inverse persistence ($\theta$), which shows positive standardized anomaly over the British Isles.

\begin{figure}[h!]
\centering
\noindent\includegraphics[width=0.95\textwidth]{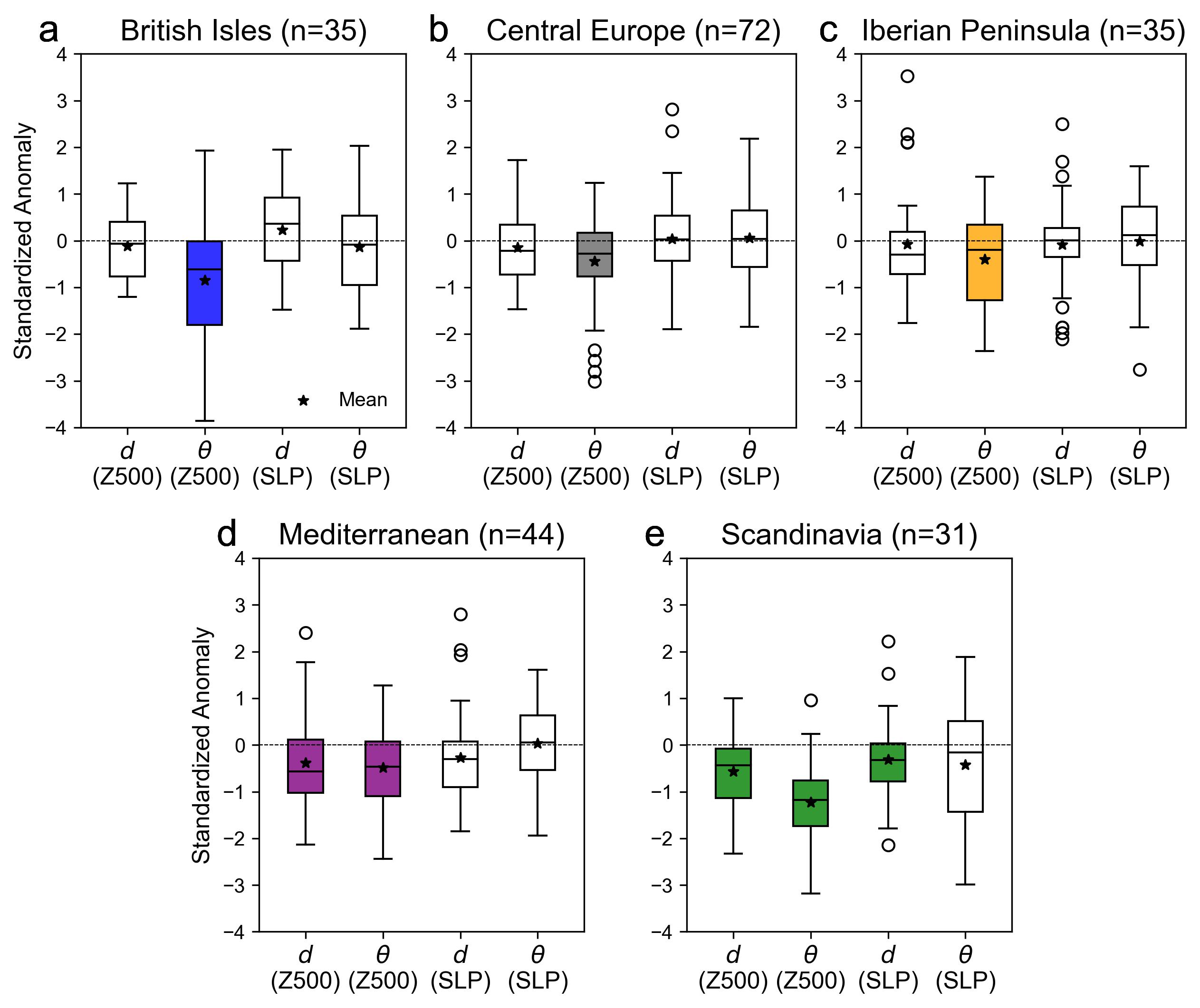}
\caption{\textbf{Box plots of standardized anomalies of dynamical indices for the dashed bounding boxes shown in Fig.~\ref{fig:fig2} using the single-domain approach.} The dynamical indices are computed within the dashed bounding boxes shown in the row 4 of Fig.~\ref{fig:fig2}, followed by standardization steps as described above. Each subplot shows the results for one studied region, with statistically significant results highlighted in the color representing that region, as in Fig.~\ref{fig:fig1}.}
\label{fig:fig5}
\end{figure}

These results are consistent with Holmberg et al.\cite{holmberg2023link}, showing that summertime heatwaves in Europe are not systematically linked with more persistent large-scale circulation patterns.
We do not consider this to be in conflict with the findings from our new approach (Fig.~\ref{fig:fig2}); rather, it underscores the importance of incorporating different spatial scales in the analysis.
Our new approach can pinpoint the regions that exhibit the most dynamical relevance of the studied phenomena, a result that was not possible to achieve with the pre-selected fixed-domain approach.
Indeed, based on the results shown in Fig.~\ref{fig:fig2}, we can identify bounding boxes where the mid-level atmosphere demonstrates significantly higher persistence (dashed bounding boxes in row 4, Fig.~\ref{fig:fig2}).
Different bounding boxes are used for heatwaves in different affected regions, as expected due to the assumption that events in different regions may have diverse atmospheric drivers.
We compute the dynamical indices of atmospheric circulation patterns within each bounding box, and plot the standardized anomalies as box plots for the heatwave onset days in the corresponding regions (see Fig.~\ref{fig:fig5}).
Consistent with expectations, more statistically significant results are observed when smaller bounding boxes are used.
In particular, significantly more persistent characteristics are displayed at the mid-level troposphere during the onset days of heatwaves for all regions.
Similarly, negative anomalies in local dimension ($d$) are observed across all regions, although they are statistically significant only in the Mediterranean and Scandinavia.
At the surface level, only the local dimension ($d$) in Scandinavia exhibits a significant negative anomaly, while the others are not significant.

\section{Discussions and conclusions}
\label{sec:discussion}

By collectively examining the results discussed above, we can recognize the importance of spatial scales in studies based on dynamical indices.
Although European heatwaves are not systematically linked with large-scale persistent atmospheric circulation patterns, they are accompanied by significant localized dynamical characteristics, as revealed by the new methodology introduced in this study.
Furthermore, we demonstrate that the guidance provided by this method can be combined with the fixed-domain approach to select bounding boxes better suited to the studied phenomena.

Dynamical systems theory has been extensively applied to diverse complex physical systems across various spatial scales~\cite{faranda2023statistical}. 
However, when addressing spatio-temporal systems, past studies have computed the metrics on a predefined bounding box based on expert knowledge. 
While sensitivity analyses on the boundaries of such bounding boxes are often discussed, the influence of the bounding box's size -- namely, the considered spatial scale -- is rarely examined.
To address these constraints, we propose a novel framework for applying dynamical indices to spatio-temporal complex systems, inspired by the concept of convolutional neural networks.
By employing a sliding bounding box with a customizable size, we extract the system's dynamical properties that are `local' in both time and space at the spatial scale of interest.
Compared to the previous approach with only two indices as outputs, this method generates spatio-temporal maps that allow for a more detailed analysis, providing richer insights into the dynamical features of certain phenomena of interest. 

Despite these advantages, certain caveats of this framework should be acknowledged.
Notably, it is significantly more computationally expensive than previous approach, and this cost cannot be mitigated.
Additionally, the choice of sliding window size relies on prior knowledge of the system. 
In many cases, obtaining meaningful insights requires analyzing results from different sliding window sizes, further increasing computational costs.
To address this, it is advisable to start with a relatively coarse version of SDIs (i.e., a large stride size) to manage computational costs, while noting that the computation can be parallelized to further enhance efficiency.
Meanwhile, our framework inherits an inherent limitation of dynamical indices: their computation relies on assumptions that the available data may not fully satisfy, and their impact cannot be quantified~\cite{datseris2023estimating}.

In this work, we applied this new framework to high-impact European summertime heatwaves to demonstrate its effectiveness.
Our results offer an explanation that reconciles the previously noted discrepancy in the persistence of heatwaves-associated circulation patterns between the meteorological view and the dynamical systems theory perspective~\cite{holmberg2023link}.
In fact, the conclusions drawn from both perspectives are valid within their respective contexts, where the glue between the two is provided by the varying dynamical characteristics of heatwaves across different spatial scales. 
Although heatwaves are not systematically linked to large-scale circulation patterns with distinct dynamical properties, they often display significant persistence and low-dimensional signatures in the regions upstream of the affected areas.
Additionally, looking at the temporal evolution of the anomaly fields of the dynamical indicators and of the meteorological fields themselves we find further confirmation of the validity of the notion of dynamical typicality of extreme events~\cite{Galfi2021,Lucarini2023}.

More broadly, we extended our analysis to summertime blocking events that frequently cause heatwaves in Europe, a phenomenon considered highly persistent by definition but, according to some recent studies employing various methodologies -- including dynamical systems theory~\cite{faranda2017dynamical,faranda2016switching}, mathematical models~\cite{lucarini2020new}, and ensemble forecasts~\cite{ferranti2015flow} -- is regarded as a relatively unpredictable configuration.
Similar to the findings for heatwaves, we observed that blocking centers exhibit significantly greater persistence and predictability, while being accompanied by configurations with contrasting dynamical properties in other regions within the Euro-Atlantic sector (see also Springer et al.~\cite{springer2024unsupervised}).
This indicates that the dynamical properties of blocking are closely tied to spatial scales, and analyses at different spatial scales have resulted in divergent conclusions in previous studies~\cite{legras1985persistent,faranda2017dynamical}.
Thus, we believe that our approach bridges the perspective of dynamical systems theory and more conventional meteorological views, providing an explanation for the divergent result present in the literature.

As a novel framework, we anticipate its broad application beyond the studied variables and domain, and even beyond atmospheric systems, to other spatio-temporal complex systems where spatial scales play a critical role.
For instance, it could be utilized to study energy cascades~\cite{leonard1975energy} in turbulent flow to better understand the transfer of energy across different scales.
In atmospheric science, its general nature allows it to be applied to any region or variable in atmospheric dynamics, particularly in the tropics, where the interplay of complex internal variabilities makes scale considerations especially crucial for analysis~\cite{dong2024indo}.
Meanwhile, the temporal evolution of dynamical indices, which has been studied in the life cycles of weather regimes~\cite{lee2024dynamical,hochman2021atlantic} and extreme events~\cite{hochman2022dynamics}, can be further complemented by our framework to provide additional insights.
Moreover, it could potentially be used to identify distant dynamical precursors of the studied phenomena -- another aspect that cannot be achieved using the fixed-domain approach.
This holds great significance for advancing our understanding of their onset mechanisms and potentially enhancing predictive modeling.

\begin{figure}[h!]
\noindent\includegraphics[width=1\textwidth]{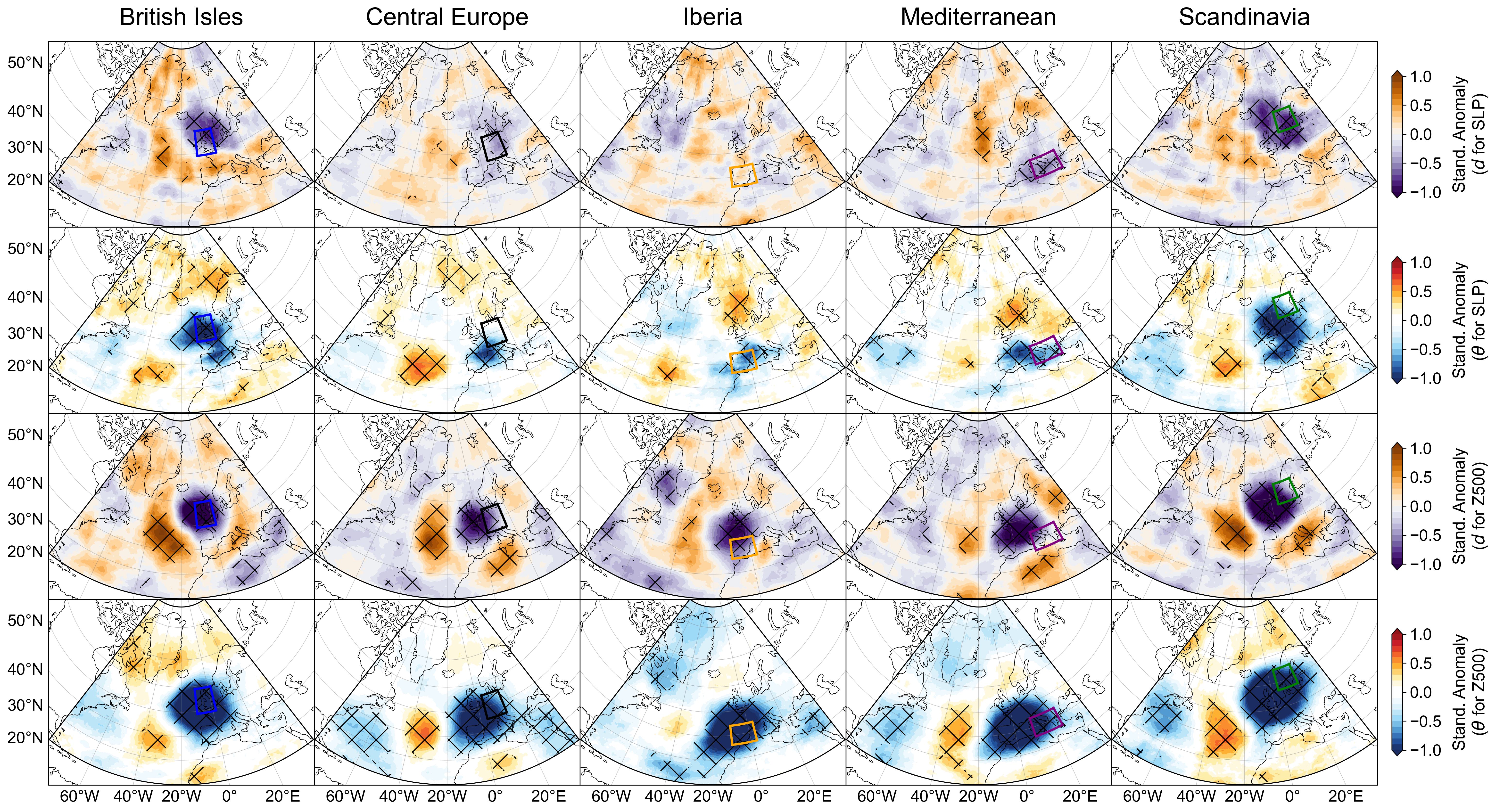}
\caption{\textbf{Composite of standardized anomalies of dynamical indices for the onset days of European summertime heatwaves.} As in Fig.~2, but for dynamical indices based on smaller window size (20° longitude by 10° latitude).}
\label{fig:smallscale}
\end{figure}

\begin{figure}[h!]
\noindent\includegraphics[width=1\textwidth]{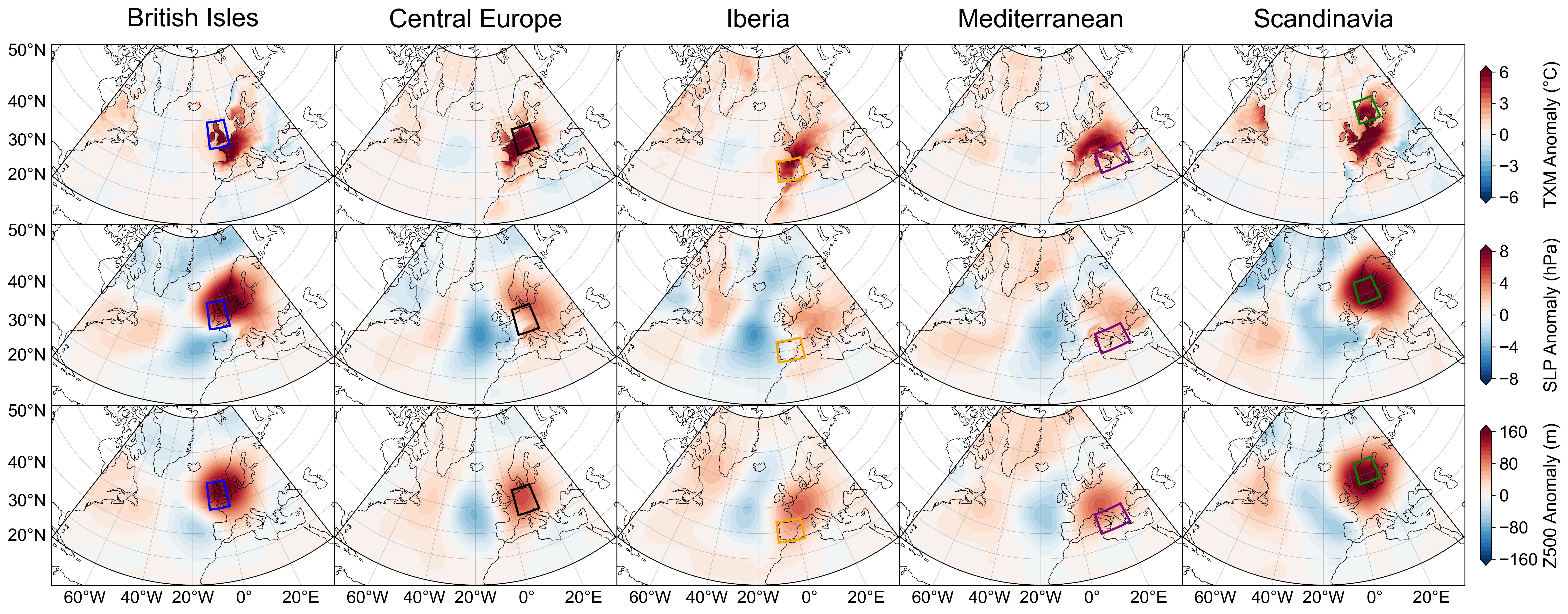}
\caption{\textbf{Onset patterns of European summertime heatwaves.} Composite anomalies for the onset days of summertime heatwaves across Europe: daily maximum 2-meter temperature (row 1), sea level pressure (row 2), and 500 hPa geopotential height (row 3).}
\label{fig:rawvariables}
\end{figure}

\begin{figure}[h!]
\noindent\includegraphics[width=1\textwidth]{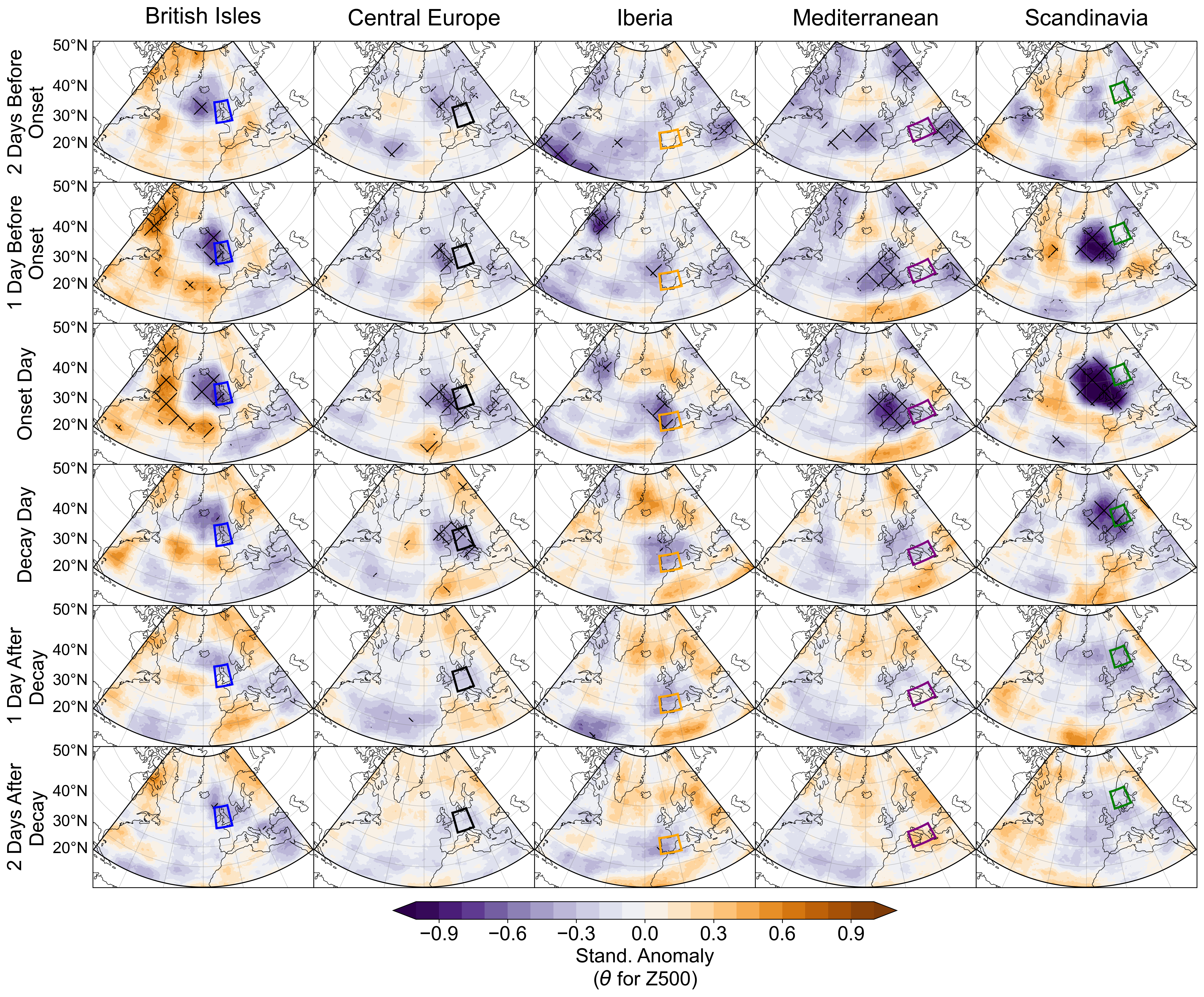}
\caption{\textbf{Temporal evolution of the composite anomaly of Z500-based local dimension ($d$) before the onset and after the decay of European summertime heatwaves.} As in Fig.~\ref{fig:fig3}, but for Z500-based local dimension.}
\label{fig:d_evolution}
\end{figure}

\begin{figure}[h!]
\noindent\includegraphics[width=1\textwidth]{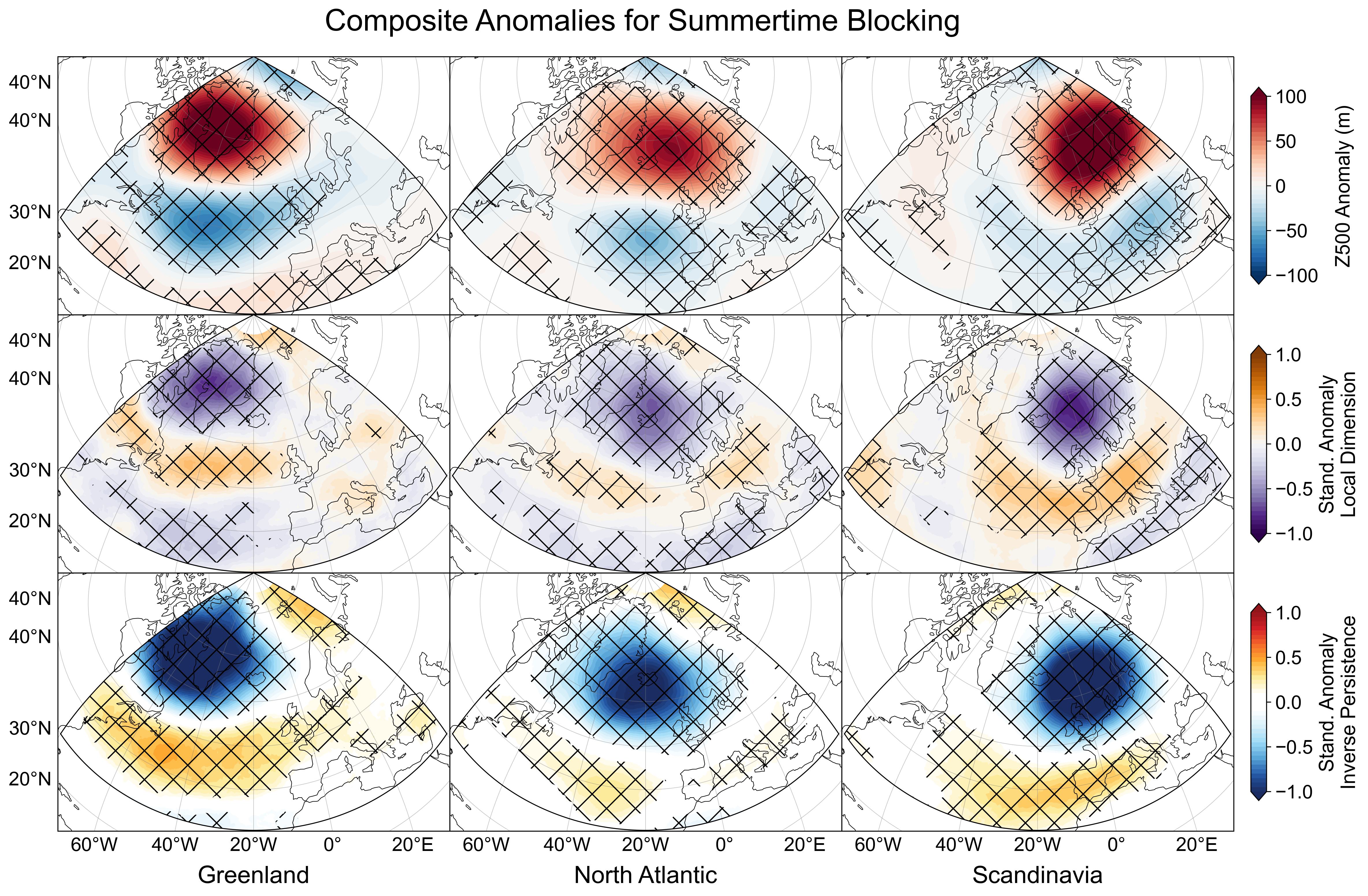}
\caption{\textbf{Composite anomalies for atmospheric blocking events.} Composite anomaly of the 500 hPa geopotential height for atmospheric blocking at different locations in the Euro-Atlantic sector (row 1). Composite of standardized anomalies for the local dimension $d$ (row 2) and the inverse persistence $\theta$ (row 3) that were computed using 500 hPa geopotential height.}
\label{fig:blocking}
\end{figure}

\clearpage
\appendix
\section{Dynamical Indices}

In this section of the supplementary text, we expand on the definition of instantaneous dimension ($d$) and inverse persistence ($\theta$) used in the main text. As introduced in the main text, these two dynamical indices grounded at the intersection of dynamical systems theory and extreme value theory, and were developed by Lucarini et al.~\cite{lucarini2016extremes} and Faranda et al.~\cite{faranda2017dynamical} to characterize local dynamical properties of complex systems. For one given state of interest $\boldsymbol{\zeta}$ in the phase space, this methodology uses its neighboring states, also known as \textit{recurrences}, to derive its state-dependent dynamical properties.
Specifically, for the considered state $\boldsymbol{\zeta}$, we first calculate its negative logarithmic distances to other states: 
\begin{equation}\label{eq:log_dist}
g\left({\boldsymbol{\zeta}}, {\bf{x}}(t)\right) = -\log \left[\operatorname{dist}\left({\bf{x}}(t), {\boldsymbol{\zeta}}\right)\right],
\end{equation}
where ${\bf{x}}(t)$ is the trajectory of the studied system and the $dist$ function can be any distance metric. Requiring a trajectory falls within a neighborhood of $\boldsymbol{\zeta}$ (a hypersphere in the phase space) is a synonym of having the time series $g[{\boldsymbol{\zeta}}, {\bf{x}}(t)]$ above one threshold $s(q, {\boldsymbol{\zeta}})$, where $s(q, {\boldsymbol{\zeta}})$ is a high threshold associated with a quantile $q$ of the series ${\bf{X}} \equiv g({\boldsymbol{\zeta}}, {\bf{x}}(t))$ ($q$ is chosen as a high value to identify neighboring states). 
Then, we define a quantity called \textit{exceedance} for the neighboring states (i.e., recurrences) of $\boldsymbol{\zeta}$ as follows:
\begin{equation}\label{eq:exceedance}
{\bf{u}}({\boldsymbol{\zeta}}) = g({\boldsymbol{\zeta}}, {\bf{x}}(t)) - s(q, {\boldsymbol{\zeta}}), \;\;\;\; \forall ~g({\boldsymbol{\zeta}}, {\bf{x}}(t)) > s(q, {\boldsymbol{\zeta}})
\end{equation}
According to extreme value theory, under the assumption of independent exceedances, the cumulative probability distribution \( F({\bf{u}}, {\boldsymbol{\zeta}}) \) follows the exponential form of the generalized Pareto distribution (GPD), expressed as:  
\begin{equation}\label{eq:cumulative}
F({\bf{u}}, {\boldsymbol{\zeta}}) \sim \exp\left[-\frac{{\bf{u}}({\boldsymbol{\zeta}})}{\sigma({\boldsymbol{\zeta}})}\right].
\end{equation}
The scale parameter $\sigma({\boldsymbol{\zeta}})$ of the distribution depends on the state ${\boldsymbol{\zeta}}$ and can be utilized to compute the local dimension as 
$d({\boldsymbol{\zeta}}) = 1/\sigma({\boldsymbol{\zeta}})$, in accordance to its definition.
The local dimension $d$ serves as an indicator of the number of active degrees of freedom in the system in the vicinity of ${\boldsymbol{\zeta}}$. Consequently, it captures the complexity of the dynamics near the state ${\boldsymbol{\zeta}}$: a system with a greater number of active degrees of freedom exhibits higher complexity.

For the other local index, inverse persistence $\theta$, its definition is adopted from the extremal index in extreme value theory. This dimensionless parameter quantifies the inverse of the clustering duration of extremes. In this context, extremes are defined as recurrences within the neighborhood of the reference state ${\boldsymbol{\zeta}}$. Consequently, persistence ($\theta^{-1}$) represents the average number of consecutive recurrences, thereby providing insight into the system's instantaneous persistence. 
Unlike the local dimension ($d$), which requires \textit{exceedances} for its estimation, the inverse persistence ($\theta$) is derived solely from the chronological order of \textit{recurrences} of the reference state ${\boldsymbol{\zeta}}$.
The full mathematical derivation of these two indices can be found in Faranda et al.~\cite{faranda2023statistical}.

\section{Definition of European Regional Heatwaves}

We define regional warm-temperature extremes over the British Isles, Scandinavia, central Europe (Germany), Iberia, and the central Mediterranean for extended summer season (JJAS), following the bounding boxes used in Zschenderlein et al.~\cite{zschenderlein2019processes}. 

To this end, we first define two criteria for each individual grid point using ERA5 daily maximum 2-meter temperature data from 1979 to 2022~\cite{russo2015top}. The first criterion requires the daily maximum temperature to exceed a threshold, defined as the 90th percentile of the daily maximum temperature for that calendar day, calculated using a centered 15-day window. The second criterion requires $M_d > 0$, where $M_d$ is a daily heatwave magnitude index defined as: 
\begin{equation}
M_{\mathrm{d}} = 
\begin{cases} 
\frac{T_{\mathrm{d}} - T_{30\mathrm{y},25\mathrm{p}}}{T_{30\mathrm{y},75\mathrm{p}} - T_{30\mathrm{y},25\mathrm{p}}}, & \text{if } T_{\mathrm{d}} > T_{30\mathrm{y},25\mathrm{p}}, \\ 
0, & \text{if } T_{\mathrm{d}} \leq T_{30\mathrm{y},25\mathrm{p}}.
\end{cases}
\end{equation}
with $T_{\mathrm{d}}$ being the daily maximum 2-meter temperature data and $T_{30\mathrm{y},75\mathrm{p}}$ ($T_{30\mathrm{y},25\mathrm{p}}$) denotes the 75th (25th) percentile of annual maximum temperatures within a 30-years sliding window. Subsequently, we aim to exclude events that are either very localized or short-lived. Accordingly, we define heatwave events as those in which at least 5\% of the predefined region (see solid boxes in Figure 2) simultaneously meet the two criteria outlined above and persist for a minimum of three consecutive days. Heatwave onset days used in this study are then identified as the first days of these events.

\section{Definition of Atmospheric Blocking events}

In this study, we primarily focus on the dynamical properties of atmospheric circulation patterns associated with European summertime heatwaves from a dynamical systems theory perspective. However, we note that another discrepancy in existing studies lies in the characterization of atmospheric blocking. While the traditional meteorological perspective considers it a highly persistent and stable configuration~\cite{legras1985persistent}, conclusions derived from dynamical systems theory suggest otherwise~\cite{faranda2017dynamical}.

We aim to address this discrepancy using analysis based on our novel framework. Nonetheless, acknowledging that heatwaves are not systematically associated with blocking events~\cite{kautz2021atmospheric}, such a conclusion cannot be drawn from the results presented in the main text. Therefore, we employ a blocking detection algorithm to rigorously define these events and, based on this, attempt to reconcile the aforementioned discrepancy.

We applied a standard blocking detection algorithm based on the reversal of 500 hPa geopotential height gradients, as described by Brunner et al.~\cite{brunner2018dependence} and references therein. For each grid point, we first compute its geopotential height gradients to the north ($\Delta Z_{\mathrm{N}}$) and to the south ($\Delta Z_{\mathrm{S}}$):
\begin{equation}
\begin{aligned}
\Delta Z_{\mathrm{N}} & =\frac{Z(\lambda, \phi+\Delta \phi)-Z(\lambda, \phi)}{\Delta \phi} \\
\Delta Z_{\mathrm{S}} & =\frac{Z(\lambda, \phi)-Z(\lambda, \phi-\Delta \phi)}{\Delta \phi}
\end{aligned}
\end{equation}
where $Z$ represents geopotential height at 500 hPa, $\Delta \phi=15^\circ$, and $\phi$ ranges from $50^{\circ}N$ to $75^{\circ}N$. Instantaneous blocking (IB) is defined at a specific grid point if the gradients simultaneously satisfy the following conditions:
\begin{equation}
\Delta Z_{\mathrm{N}} < -10m/(^{\circ}lat); \Delta Z_{\mathrm{S}} > 0m/(^{\circ}lat)
\end{equation}
Next, we apply spatiotemporal filtering to the IB field to extract large-scale and slow-moving events. The maximum IB index within ±$4^{\circ}$ latitude is taken to account for meridional movement. Extended IB cases are then selected if they span at least $15^{\circ}$ longitude, filtering out systems that are too small. Finally, blocking is defined as an extended IB persisting within ±$10^{\circ}$ longitude for at least five consecutive days, ensuring the detection of only persistent and slow-moving systems.

Blocked days are further determined for three 30° longitude regions, namely Greenland ($60^{\circ}$W to $30^{\circ}$W), North Atlantic ($30^{\circ}$W to $0^{\circ}$), and Scandinavian ($0^{\circ}$ to $30^{\circ}$E). A day is considered blocked in a given region if a block spans more than half of the region (i.e., exceeds 15° of longitude within the region). Based on this, we can plot the composite anomalies of the standardized anomalies for blocked days within the three regions, as shown in Fig.~\ref{fig:blocking}.






\clearpage
\bibliographystyle{elsarticle-num} 
\bibliography{bibliography}

\begin{thebibliography}{10}
\expandafter\ifx\csname url\endcsname\relax
  \def\url#1{\texttt{#1}}\fi
\expandafter\ifx\csname urlprefix\endcsname\relax\def\urlprefix{URL }\fi
\expandafter\ifx\csname href\endcsname\relax
  \def\href#1#2{#2} \def\path#1{#1}\fi

\bibitem{lorenz1963deterministic}
E.~N. Lorenz, Deterministic nonperiodic flow, Journal of atmospheric sciences 20~(2) (1963) 130--141.

\bibitem{altman2018curse}
N.~Altman, M.~Krzywinski, The curse (s) of dimensionality, Nat Methods 15~(6) (2018) 399--400.

\bibitem{gao2024similarity}
X.-T. Gao, B.~Tian, Similarity reductions on a (2+ 1)-dimensional variable-coefficient modified kadomtsev-petviashvili system describing certain electromagnetic waves in a thin film, International Journal of Theoretical Physics 63~(4) (2024) 99.

\bibitem{gao2025inhomogeneity}
X.-Y. Gao, J.-G. Liu, G.-W. Wang, Inhomogeneity, magnetic auto-b{\"a}cklund transformations and magnetic solitons for a generalized variable-coefficient kraenkel-manna-merle system in a deformed ferrite, Applied Mathematics Letters (2025) 109615.

\bibitem{grassberger1986climatic}
P.~Grassberger, Do climatic attractors exist?, Nature 323~(6089) (1986) 609--612.

\bibitem{lorenz1991dimension}
E.~N. Lorenz, Dimension of weather and climate attractors, Nature 353~(6341) (1991) 241--244.

\bibitem{lucarini2016extremes}
V.~Lucarini, D.~Faranda, J.~M.~M. de~Freitas, M.~Holland, T.~Kuna, M.~Nicol, M.~Todd, S.~Vaienti, et~al., Extremes and recurrence in dynamical systems, John Wiley \& Sons, 2016.

\bibitem{faranda2017dynamical}
D.~Faranda, G.~Messori, P.~Yiou, Dynamical proxies of north atlantic predictability and extremes, Scientific reports 7~(1) (2017) 41278.

\bibitem{hochman2022dynamics}
A.~Hochman, S.~Scher, J.~Quinting, J.~G. Pinto, G.~Messori, Dynamics and predictability of cold spells over the eastern mediterranean, Climate Dynamics 58~(7) (2022) 2047--2064.

\bibitem{holmberg2023link}
E.~Holmberg, G.~Messori, R.~Caballero, D.~Faranda, The link between european warm-temperature extremes and atmospheric persistence, Earth System Dynamics 14~(4) (2023) 737--765.

\bibitem{messori2017dynamical}
G.~Messori, R.~Caballero, D.~Faranda, A dynamical systems approach to studying midlatitude weather extremes, Geophysical Research Letters 44~(7) (2017) 3346--3354.

\bibitem{hochman2021atlantic}
A.~Hochman, G.~Messori, J.~F. Quinting, J.~G. Pinto, C.~M. Grams, Do atlantic-european weather regimes physically exist?, Geophysical Research Letters 51~(14) (2021).

\bibitem{lee2024dynamical}
S.~H. Lee, G.~Messori, The dynamical footprint of year-round north american weather regimes, Geophysical Research Letters 51~(2) (2024) e2023GL107161.

\bibitem{platzer2021dynamical}
P.~Platzer, B.~Chapron, P.~Tandeo, Dynamical properties of weather regime transitions, in: Stochastic Transport in Upper Ocean Dynamics Annual Workshop, Springer, 2021, pp. 223--236.

\bibitem{liu2021dynamical}
G.~Liu, F.~Falasca, A.~Bracco, Dynamical characterization of the loop current attractor, Geophysical Research Letters 48~(24) (2021) e2021GL096731.

\bibitem{gualandi2024similarities}
A.~Gualandi, L.~Dal~Zilio, D.~Faranda, G.~Mengaldo, Similarities and differences between natural and simulated slow earthquakes, Geophysical Research Letters 48~(20) (2024) e2021GL095574.

\bibitem{springer2024unsupervised}
S.~Springer, A.~Laio, V.~M. Galfi, V.~Lucarini, Unsupervised detection of large-scale weather patterns in the northern hemisphere via markov state modelling: from blockings to teleconnections, npj Climate and Atmospheric Science 7~(1) (2024) 105.

\bibitem{alberti2021small}
T.~Alberti, D.~Faranda, R.~V. Donner, T.~Caby, V.~Carbone, G.~Consolini, B.~Dubrulle, S.~Vaienti, Small-scale induced large-scale transitions in solar wind magnetic field, The Astrophysical journal letters 914~(1) (2021) L6.

\bibitem{alberti2023scale}
T.~Alberti, D.~Faranda, V.~Lucarini, R.~V. Donner, B.~Dubrulle, F.~Daviaud, Scale dependence of fractal dimension in deterministic and stochastic lorenz-63 systems, Chaos: An Interdisciplinary Journal of Nonlinear Science 33~(2) (2023).

\bibitem{lv2016multivariate}
Y.~Lv, R.~Yuan, G.~Song, Multivariate empirical mode decomposition and its application to fault diagnosis of rolling bearing, Mechanical Systems and Signal Processing 81 (2016) 219--234.

\bibitem{slonosky2002does}
V.~Slonosky, P.~Yiou, Does the nao index represent zonal flow? the influence of the nao on north atlantic surface temperature, Climate Dynamics 19~(1) (2002) 17--30.

\bibitem{miralles2014mega}
D.~G. Miralles, A.~J. Teuling, C.~C. Van~Heerwaarden, J.~Vil{\`a}-Guerau~de Arellano, Mega-heatwave temperatures due to combined soil desiccation and atmospheric heat accumulation, Nature geoscience 7~(5) (2014) 345--349.

\bibitem{zschenderlein2019processes}
P.~Zschenderlein, A.~H. Fink, S.~Pfahl, H.~Wernli, Processes determining heat waves across different european climates, Quarterly Journal of the Royal Meteorological Society 145~(724) (2019) 2973--2989.

\bibitem{russo2015top}
S.~Russo, J.~Sillmann, E.~M. Fischer, Top ten european heatwaves since 1950 and their occurrence in the coming decades, Environmental Research Letters 10~(12) (2015) 124003.

\bibitem{stefanon2012heatwave}
M.~Stefanon, F.~D’Andrea, P.~Drobinski, Heatwave classification over europe and the mediterranean region, Environmental Research Letters 7~(1) (2012) 014023.

\bibitem{hochman2019new}
A.~Hochman, P.~Alpert, T.~Harpaz, H.~Saaroni, G.~Messori, A new dynamical systems perspective on atmospheric predictability: Eastern mediterranean weather regimes as a case study, Science advances 5~(6) (2019) eaau0936.

\bibitem{dong2024revisiting}
C.~Dong, D.~Faranda, A.~Gualandi, V.~Lucarini, G.~Mengaldo, Revisiting the predictability of dynamical systems: a new local data-driven approach, arXiv preprint arXiv:2409.14865 (2024).

\bibitem{faranda2023statistical}
D.~Faranda, G.~Messori, T.~Alberti, C.~Alvarez-Castro, T.~Caby, L.~Cavicchia, E.~Coppola, R.~Donner, B.~Dubrulle, V.~M. Galfi, et~al., A statistical physics and dynamical systems perspective on geophysical extreme events, arXiv preprint arXiv:2309.15393 (2023).

\bibitem{faranda2023dynamical}
D.~Faranda, G.~Messori, P.~Yiou, S.~Thao, F.~Pons, B.~Dubrulle, Dynamical footprints of hurricanes in the tropical dynamics, Chaos: An Interdisciplinary Journal of Nonlinear Science 33~(1) (2023).

\bibitem{o2015introduction}
K.~O'shea, R.~Nash, An introduction to convolutional neural networks, arXiv preprint arXiv:1511.08458 (2015).

\bibitem{garcia2010review}
R.~Garc{\'\i}a-Herrera, J.~D{\'\i}az, R.~M. Trigo, J.~Luterbacher, E.~M. Fischer, A review of the european summer heat wave of 2003, Critical Reviews in Environmental Science and Technology 40~(4) (2010) 267--306.

\bibitem{gallo2024heat}
E.~Gallo, M.~Quijal-Zamorano, R.~F. M{\'e}ndez~Turrubiates, C.~Tonne, X.~Basaga{\~n}a, H.~Achebak, J.~Ballester, Heat-related mortality in europe during 2023 and the role of adaptation in protecting health, Nature medicine (2024) 1--5.

\bibitem{ebi2021extreme}
K.~L. Ebi, J.~Vanos, J.~W. Baldwin, J.~E. Bell, D.~M. Hondula, N.~A. Errett, K.~Hayes, C.~E. Reid, S.~Saha, J.~Spector, et~al., Extreme weather and climate change: population health and health system implications, Annual review of public health 42~(1) (2021) 293--315.

\bibitem{domeisen2023prediction}
D.~I. Domeisen, E.~A. Eltahir, E.~M. Fischer, R.~Knutti, S.~E. Perkins-Kirkpatrick, C.~Sch{\"a}r, S.~I. Seneviratne, A.~Weisheimer, H.~Wernli, Prediction and projection of heatwaves, Nature Reviews Earth \& Environment 4~(1) (2023) 36--50.

\bibitem{pfahl2012quantifying}
S.~Pfahl, H.~Wernli, Quantifying the relevance of atmospheric blocking for co-located temperature extremes in the northern hemisphere on (sub-) daily time scales, Geophysical Research Letters 39~(12) (2012).

\bibitem{cassou2005tropical}
C.~Cassou, L.~Terray, A.~S. Phillips, Tropical atlantic influence on european heat waves, Journal of climate 18~(15) (2005) 2805--2811.

\bibitem{legras1985persistent}
B.~Legras, M.~Ghil, Persistent anomalies, blocking and variations in atmospheric predictability, Journal of Atmospheric Sciences 42~(5) (1985) 433--471.

\bibitem{faranda2016switching}
D.~Faranda, G.~Masato, N.~Moloney, Y.~Sato, F.~Daviaud, B.~Dubrulle, P.~Yiou, The switching between zonal and blocked mid-latitude atmospheric circulation: a dynamical system perspective, Climate Dynamics 47 (2016) 1587--1599.

\bibitem{lucarini2020new}
V.~Lucarini, A.~Gritsun, A new mathematical framework for atmospheric blocking events, Climate Dynamics 54~(1) (2020) 575--598.

\bibitem{schubert2016dynamical}
S.~Schubert, V.~Lucarini, Dynamical analysis of blocking events: spatial and temporal fluctuations of covariant lyapunov vectors, Quarterly Journal of the Royal Meteorological Society 142~(698) (2016) 2143--2158.

\bibitem{hersbach2020era5}
H.~Hersbach, B.~Bell, P.~Berrisford, S.~Hirahara, A.~Hor{\'a}nyi, J.~Mu{\~n}oz-Sabater, J.~Nicolas, C.~Peubey, R.~Radu, D.~Schepers, et~al., The era5 global reanalysis, Quarterly Journal of the Royal Meteorological Society 146~(730) (2020) 1999--2049.

\bibitem{jezequel2018role}
A.~J{\'e}z{\'e}quel, P.~Yiou, S.~Radanovics, Role of circulation in european heatwaves using flow analogues, Climate dynamics 50~(3) (2018) 1145--1159.

\bibitem{cassou2016disruption}
C.~Cassou, J.~Cattiaux, Disruption of the european climate seasonal clock in a warming world, Nature Climate Change 6~(6) (2016) 589--594.

\bibitem{Galfi2021}
V.~M. Galfi, V.~Lucarini, \href{https://link.aps.org/doi/10.1103/PhysRevLett.127.058701}{{Fingerprinting Heatwaves and Cold Spells and Assessing Their Response to Climate Change Using Large Deviation Theory}}, Phys. Rev. Lett. 127 (2021) 058701.
\newblock \href {https://doi.org/10.1103/PhysRevLett.127.058701} {\path{doi:10.1103/PhysRevLett.127.058701}}.
\newline\urlprefix\url{https://link.aps.org/doi/10.1103/PhysRevLett.127.058701}

\bibitem{Lucarini2023}
V.~Lucarini, V.~M. Galfi, J.~Riboldi, G.~Messori, \href{https://dx.doi.org/10.1088/1748-9326/acab77}{{Typicality of the 2021 Western North America summer heatwave}}, Environmental Research Letters 18~(1) (2023) 015004.
\newblock \href {https://doi.org/10.1088/1748-9326/acab77} {\path{doi:10.1088/1748-9326/acab77}}.
\newline\urlprefix\url{https://dx.doi.org/10.1088/1748-9326/acab77}

\bibitem{Speranza1983}
A.~Speranza, \href{https://doi.org/10.1007/BF02590154}{Deterministic and statistical properties of the westerlies}, pure and applied geophysics 121~(3) (1983) 511--562.
\newblock \href {https://doi.org/10.1007/BF02590154} {\path{doi:10.1007/BF02590154}}.
\newline\urlprefix\url{https://doi.org/10.1007/BF02590154}

\bibitem{DellAquila2005}
A.~Dell’Aquila, V.~Lucarini, P.~M. Ruti, S.~Calmanti, \href{https://doi.org/10.1007/s00382-005-0048-x}{Hayashi spectra of the northern hemisphere mid-latitude atmospheric variability in the ncep–ncar and ecmwf reanalyses}, Climate Dynamics 25~(6) (2005) 639--652.
\newblock \href {https://doi.org/10.1007/s00382-005-0048-x} {\path{doi:10.1007/s00382-005-0048-x}}.
\newline\urlprefix\url{https://doi.org/10.1007/s00382-005-0048-x}

\bibitem{Lucarini2007}
V.~Lucarini, S.~Calmanti, A.~Dell’Aquila, P.~M. Ruti, A.~Speranza, \href{https://doi.org/10.1007/s00382-006-0213-x}{Intercomparison of the northern hemisphere winter mid-latitude atmospheric variability of the ipcc models}, Climate Dynamics 28~(7) (2007) 829--848.
\newblock \href {https://doi.org/10.1007/s00382-006-0213-x} {\path{doi:10.1007/s00382-006-0213-x}}.
\newline\urlprefix\url{https://doi.org/10.1007/s00382-006-0213-x}

\bibitem{fery2022learning}
L.~Fery, B.~Dubrulle, B.~Podvin, F.~Pons, D.~Faranda, Learning a weather dictionary of atmospheric patterns using latent dirichlet allocation, Geophysical Research Letters 49~(9) (2022) e2021GL096184.

\bibitem{wilks2016stippling}
D.~Wilks, “the stippling shows statistically significant grid points”: How research results are routinely overstated and overinterpreted, and what to do about it, Bulletin of the American Meteorological Society 97~(12) (2016) 2263--2273.

\bibitem{woollings2018blocking}
T.~Woollings, D.~Barriopedro, J.~Methven, S.-W. Son, O.~Martius, B.~Harvey, J.~Sillmann, A.~R. Lupo, S.~Seneviratne, Blocking and its response to climate change, Current climate change reports 4 (2018) 287--300.

\bibitem{kautz2021atmospheric}
L.-A. Kautz, O.~Martius, S.~Pfahl, J.~G. Pinto, A.~M. Ramos, P.~M. Sousa, T.~Woollings, Atmospheric blocking and weather extremes over the euro-atlantic sector--a review, Weather and Climate Dynamics Discussions 2021 (2021) 1--43.

\bibitem{brunner2018dependence}
L.~Brunner, N.~Schaller, J.~Anstey, J.~Sillmann, A.~K. Steiner, Dependence of present and future european temperature extremes on the location of atmospheric blocking, Geophysical research letters 45~(12) (2018) 6311--6320.

\bibitem{Galfi2021RNV}
V.~M. G{\'a}lfi, V.~Lucarini, F.~Ragone, J.~Wouters, \href{https://doi.org/10.1007/s40766-021-00020-z}{Applications of large deviation theory in geophysical fluid dynamics and climate science}, La Rivista del Nuovo Cimento 44~(6) (2021) 291--363.
\newblock \href {https://doi.org/10.1007/s40766-021-00020-z} {\path{doi:10.1007/s40766-021-00020-z}}.
\newline\urlprefix\url{https://doi.org/10.1007/s40766-021-00020-z}

\bibitem{Dematteis2018}
G.~Dematteis, T.~Grafke, E.~Vanden-Eijnden, \href{https://www.pnas.org/content/115/5/855}{Rogue waves and large deviations in deep sea}, Proceedings of the National Academy of Sciences 115~(5) (2018) 855--860.
\newblock \href {http://arxiv.org/abs/https://www.pnas.org/content/115/5/855.full.pdf} {\path{arXiv:https://www.pnas.org/content/115/5/855.full.pdf}}, \href {https://doi.org/10.1073/pnas.1710670115} {\path{doi:10.1073/pnas.1710670115}}.
\newline\urlprefix\url{https://www.pnas.org/content/115/5/855}

\bibitem{datseris2023estimating}
G.~Datseris, I.~Kottlarz, A.~P. Braun, U.~Parlitz, Estimating fractal dimensions: A comparative review and open source implementations, Chaos: An Interdisciplinary Journal of Nonlinear Science 33~(10) (2023).

\bibitem{ferranti2015flow}
L.~Ferranti, S.~Corti, M.~Janousek, Flow-dependent verification of the ecmwf ensemble over the euro-atlantic sector, Quarterly Journal of the Royal Meteorological Society 141~(688) (2015) 916--924.

\bibitem{leonard1975energy}
A.~Leonard, Energy cascade in large-eddy simulations of turbulent fluid flows, in: Advances in geophysics, Vol.~18, Elsevier, 1975, pp. 237--248.

\bibitem{dong2024indo}
C.~Dong, R.~Noyelle, G.~Messori, A.~Gualandi, L.~Fery, P.~Yiou, M.~Vrac, F.~D'Andrea, S.~J. Camargo, E.~Coppola, G.~Balsamo, C.~Chen, D.~Faranda, G.~Mengaldo, Indo-pacific regional extremes aggravated by changes in tropical weather patterns, Nature Geoscience (2024).

\end{thebibliography}
\end{document}